\documentclass{webofc}

\usepackage[varg]{txfonts}   
\usepackage{hyperref}
\usepackage{url}
\usepackage{graphicx}
\usepackage{caption}
\hypersetup{colorlinks=true,citecolor=blue,urlcolor=blue,linkcolor=blue}
\begin{document}
\title{$4b + X$ via electroweak multi-Higgs production as smoking gun signals for Type-I 2HDM at the LHC}
%
%

\author{\firstname{Prasenjit} \lastname{Sanyal}\inst{1}\fnsep\thanks{\email{prasenjit.sanyal01@gmail.com}} \and
        \firstname{Tanmoy} \lastname{Mondal}\inst{2}\fnsep\thanks{\email{tanmoy.mondal@pilani.bits-pilani.ac.in}} \and
        \firstname{Stefano} \lastname{Moretti}\inst{3,4}\fnsep\thanks{\email{s.moretti@soton.ac.uk,  stefano.moretti@physics.uu.se}} \and
        \firstname{Shoaib} \lastname{Munir}\inst{5}\fnsep\thanks{\email{munir1@stolaf.edu}}
}

\institute{Department of Physics, Konkuk University, Seoul 05029, Republic of Korea
\and
           Birla Institute of Technology and Science, Pilani, 333031, Rajasthan, India
\and
        School of Physics and Astronomy, University of Southampton, Southampton SO17 1BJ, United Kingdom
\and           
           Department of Physics and Astronomy, Uppsala University, Box 516, SE-751 20 Uppsala, Sweden
\and
           Department of Physics, Faculty of Natural Sciences and Mathematics, St. Olaf College, Northfield, MN 55057, United States                 
          }

\abstract{Extending the Standard Model (SM) by one additional Higgs doublet leads to the Two-Higgs Doublet Model (2HDM). A specific charge assignment of the SM fermions under the $\mathbb{Z}_2$ symmetry leads to the Type-I 2HDM. A key feature of the Type-I 2HDM is that all the additional Higgs bosons can be fermiophobic, when their couplings to the SM fermions are suppressed. As a result, all the new Higgs states can be fairly light, $\sim$100 GeV or less, without being in conflict with the current data from the direct Higgs boson searches and the $B$-physics measurements. In a recent study Ref.~\cite{Mondal:2023wib}, which this proceeding is based on, we established that the new neutral as well as the charged Higgs bosons in this model can all be simultaneously observable in the multi-$b$ final state. An experimental validation of our results would be a clear indication that the true underlying Higgs sector in nature is the Type-I 2HDM.}
\maketitle
\section{Introduction}
\label{intro}  The minimal Electroweak Symmetry Breaking (EWSB) dynamics provided by the SM of particle physics have been extensively tested, and have so far been in very good agreement with the results from many collider experiments. The discovery of a neutral Higgs boson with a mass of 125 GeV strongly favors the Higgs mechanism of EWSB in the SM, further raising its status as the only acceptable framework at the energy scale accessible at the present day colliders. However, the SM is still widely believed to be an incomplete framework, because of several physical phenomena it is unable to explain. Observation of additional Higgs boson will therefore provide a strong indication that the actual EWSB dynamics are not of a minimal nature.

The 2HDM is one of the most minimal extensions of the SM, and contains an additional Higgs doublet.
After the EWSB, the scalar sector in the CP-conserving limit of the 2HDM consists of two CP-even scalars, $h$ and $H$, a CP-odd scalar (pseudoscalar), $A$, and a pair of charged Higgs bosons, $H^\pm$. Here we follow the standard mass hierarchy, wherein the observed 125 GeV Higgs boson is identified with the $h$, the lighter of the two scalars. The most general 2HDM, where both the Higgs doublets couple to all the fermions, suffers from the Flavor-Changing Neutral Current (FCNC) issue. To circumvent this issue, a $\mathbb{Z}_2$ symmetry is generally imposed. When $\Phi_1 \to \Phi_1$, $\Phi_2 \to -\Phi_2$, $u^i_R \to -u^i_R$, $d^i_R \to -d^i_R$, $e^i_R \to -e^i_R$ under this $\mathbb{Z}_2$ symmetry, all the quarks and charged leptons couple only to $\Phi_2$, and the resulting model is referred to as the Type-I 2HDM.
 
The conventional search strategies to look for single or multiple Higgs states at the LHC involve QCD-induced production processes like the gluon fusion or the $b\bar{b}$-annihilation, where the $b$-quarks are themselves produced from a (double) gluon-splitting. These QCD-induced productions can be subdominant in some beyond SM (BSM) frameworks if the Higgs boson couplings are not SM-like. The Type-I 2HDM fits this criterion, as all the heavier Higgs bosons can show fermiophobic behavior, when their couplings to the fermions are suppressed \cite{Enberg:2018pye, Mondal:2021bxa, Kim:2022nmm, Kim:2023lxc, Sanyal:2023pfs, Wang:2023pqx}. In this situation, the production of the BSM Higgs bosons through the EW production processes can dominate the QCD ones. Similarly, the bosonic decays of the BSM Higgs bosons can dominate their fermionic decay modes, if kinematically allowed. Moreover, the $q\bar{q}'$-induced (where $q$ represents the valence quarks $u$ and $d$) EW production of a single charged Higgs state does not have a QCD counterpart.
 
In this work, through a detector level Monte Carlo (MC) analysis, we establish that the EW production can provide simultaneous visible signals of all the BSM Higgs bosons of the Type-I 2HDM at the LHC with 3000fb$^{-1}$ luminosity. The parameter space configuration where this is possible requires $A$ to be lighter than $h$, with $H$ and $H^\pm$ also not much heavier than it. Our signature channel, constituting of multiple $b$-quarks, allows a full reconstruction of the $A$, $H$ and $H^\pm$ masses. We also briefly explore the prospects of such EW-induced multi-Higgs production at the future electron-positron colliders.

\section{The Type-I 2HDM}
\label{sec-1}

The most general potential of a CP-conserving 2HDM \cite{Branco:2011iw} can be written as
\begin{equation}
\begin{split}
\mathcal{V}(\Phi_1,\Phi_2) &= m_{11}^2\Phi_1^\dagger\Phi_1+ m_{22}^2\Phi_2^\dagger\Phi_2
-[m_{12}^2\Phi_1^\dagger\Phi_2+ \, \text{h.c.} ] +\frac{\lambda_1}{2}(\Phi_1^\dagger\Phi_1)^2
+\frac{\lambda_2}{2}(\Phi_2^\dagger\Phi_2)^2\\
&+\lambda_3(\Phi_1^\dagger\Phi_1)(\Phi_2^\dagger\Phi_2)
+\lambda_4(\Phi_1^\dagger\Phi_2)(\Phi_2^\dagger\Phi_1) 
+\left[\frac{\lambda_5}{2}(\Phi_1^\dagger\Phi_2)^2
+\text{h.c.}\right].
\label{eq:2hdmpot}
\end{split}
\end{equation}
where $\Phi_{1,2}$ are the two Higgs doublets with hypercharges $Y=\pm 1/2$. The $\mathbb{Z}_2$ symmetry is softly broken by the $m_{12}^2$ term, which is real, as are all the $\lambda$ couplings. After the EWSB, the Higgs doublets $\Phi_{1,2}$, can be represented in terms of their respective vacuum expectation values (VEVs) $v_1$ and $v_2$, the Goldstone boson $G$ and
$G^\pm$, and the physical Higgs states, as

\begin{align}
\Phi_1=\frac{1}{\sqrt{2}}\left(\begin{array}{c}
 \sqrt{2}\left(G^+ c_\beta -H^+ s_\beta\right)  \\
 v_1-h s_\alpha+H c_\alpha+i\left( G c_\beta-A s_\beta \right)
\end{array}
\right), ~~
\Phi_2=\frac{1}{\sqrt{2}}\left(\begin{array}{c}
 \sqrt{2}\left(G^+ s_\beta +H^+c_\beta\right)  \\
 v_2+h c_\alpha+Hs_\alpha+i\left( G s_\beta+A c_\beta \right)
\end{array} 
\right)\,,
\end{align}
where $v=\sqrt{v_1^2+v_2^2}=246$ GeV, and we have defined the parameter $\tan\beta = v_2/v_1$. We use the abbreviations $s_x$ ($c_x$) to denote $\sin(x)$ ($\cos(x)$), and $t_{\beta}$ for $\tan\beta$. The physical state $h$ is identified as the SM-like observed Higgs boson with mass fixed to 125 GeV and almost exactly SM-like couplings, which can be achieved in the alignment limit, $s_{\beta - \alpha} \to 1$. 

The Yukawa Lagrangian in the Type-I 2HDM can be written in terms of the physical Higgs states as
\begin{equation}
\begin{split}
\mathcal{L}_\text{Yuk} &=
-\sum_{f=u,d,\ell} \frac{m_f}{v}\left(\xi_h^f\bar{f}fh +
\xi_H^f\bar{f}fH - i\xi_A^f\bar{f}f\gamma_5A \right) \\ 
&-\Big\lbrace \frac{\sqrt{2}V_{ud}}{v}\bar{u}\,\Big(\xi^u_A m_u P_L + \xi^d_A m_d P_R\Big)\,dH^+ 
+\frac{\sqrt{2}m_\ell}{v} \xi^\ell_A\bar{\nu}_L \ell_R H^+ + \text{h.c.} \Big\rbrace \, .
\end{split}
\end{equation}
where $V$ is the CKM matrix, and $P_{L,R} = \frac{1}{2}(1\mp\gamma_5)$ are the chirality projection operators. The Yukawa coupling modifiers $\xi^f$ are given in Tab.~\ref{Tab:YukawaFactors}.
\begin{table}[t]
\setlength{\tabcolsep}{2pt}
\begin{center}
\begin{tabular}{|c||c|c|c|c|c|c|c|c|c|}
\hline
~Type-I~& ~$\xi_h^u$~ & ~$\xi_h^d$~ & ~$\xi_h^\ell$~
& ~$\xi_H^u$~ & ~$\xi_H^d$~ & ~$\xi_H^\ell$~
& ~$\xi_A^u$~ & ~$\xi_A^d$~ & ~$\xi_A^\ell$~ \\ \cline{2-10}
~2HDM~& ~$c_\alpha/s_\beta$~ & ~$c_\alpha/s_\beta$~ & ~$c_\alpha/s_\beta$~
& ~$s_\alpha/s_\beta$~ & ~$s_\alpha/s_\beta$~ & ~$s_\alpha/s_\beta$~
& ~$1/t_\beta$~ & ~$-1/t_\beta$~ & ~$-1/t_\beta$~ \\
 \hline
\end{tabular}
\end{center}
 \caption{The Yukawa coupling modifiers in the Type-I 2HDM.}
\label{Tab:YukawaFactors}
\end{table} 

The couplings of the neutral Higgs bosons to pairs of gauge bosons ($V= Z,~W^\pm$) read
\begin{align}
g_{h{VV}} &= s_{\beta-\alpha} g_{h{VV}}^{SM}, \quad g_{H{VV}} = c_{\beta -\alpha} g_{h{VV}}^{SM}, \quad g_{A{VV}} = 0  , 
\label{hVV}
\end{align}
while the $Z$-boson couplings to the scalar-pseudoscalar pairs are given as
\begin{align}
g_{hAZ_\mu} = \frac{g}{2 c_{\theta_W}}c_{\beta-\alpha}(p_h - p_A)_\mu,\quad g_{HAZ_\mu} = -\frac{g}{2 c_{\theta_W}}s_{\beta-\alpha}(p_H - p_A)_\mu .
\label{hhZ}
\end{align}
Finally, the $W$-boson couplings to the charged Higgs boson are 
\begin{align}
g_{H^\mp W^\pm h} = \mp\frac{ig}{2}c_{\beta - \alpha}(p_h - p_{H^\mp})_\mu, 
~~ g_{H^\mp W^\pm H} = \pm\frac{ig}{2}s_{\beta-\alpha}(p_H - p_{H^\mp})_\mu,
~~ g_{H^\mp W^\pm A} = \frac{g}{2}(p_A - p_{H^\mp})_\mu \,,
\label{hhW}
\end{align}
where $p_\mu$ represents the 4-momenta of the incoming Higgs bosons. From the Tab.~\ref{Tab:YukawaFactors}, we can see that the fermionic couplings of $H^\pm$ and $A$ are suppressed at large $t_\beta$. The alignment limit, together with large $t_\beta$, is what makes the $H$ fermiophobic, which becomes more apparent by looking at its Yukawa coupling modifier,
\begin{align}
\xi^f_H = \frac{s_{\alpha}}{s_{\beta}} = c_{\beta-\alpha} - \frac{s_{\beta-\alpha}}{t_\beta}\,. 
\end{align}
For $s_{\beta-\alpha} \to 1$ and $t_{\beta} \gg 1$, $\xi^f_H$ approaches zero. The possible fermiophobic nature of the BSM Higgs bosons is one of the most striking characteristics of the Type-I 2HDM.

\section{Multi-$A$ production and benchmark scenarios}
\label{sec-2}

In Ref.~\cite{Enberg:2018pye}, it was shown that the cross section for the EW production of the $HA$ pair can be up to two orders of magnitude larger than the QCD-induced one. For parameter space regions where the $H$ has a mass above the $AA$ or $AZ$ decay thresholds, the production cross section of the subsequent states $AAA$ and/or $AAZ$ remains quite substantial. In the alignment limit, the $Hhh$ coupling is also suppressed compared to the $HAA$ coupling. Furthermore, the $HAZ$ coupling ($\propto s_{\beta-\alpha}$), and hence the BR($H\to AZ$), is also enhanced, while the $H\to VV$ decays are suppressed, even when kinematically available.

Our analysis is therefore restricted to small, $\sim$70 GeV or lower, values of $m_A$. For such a light $A$, $b\bar{b}$ is by far the dominant decay mode, even in the fermiophobic limit of the model. The multi-Higgs states that we are interested in here are thus the ones yielding at least 4 $b$-quarks via intermediate $A$s. The contributing EW process thus contain either a pair of neutral Higgs bosons, as
\begin{eqnarray}
\quad \quad\quad \quad\quad \quad\quad \quad AAA &:& q \bar{q} \to H (\to A A) A \to 4b+X\,, \nonumber \\
AAZ &:& q \bar{q} \to H (\to A Z) A \to 4b+X\,, \nonumber \\
AAWW &:& q \bar{q}\to H^+ (\to A W) H^-(\to A W)\to 4b+X\,, \nonumber
\nonumber   
\end{eqnarray}
or a pair of charged Higgs bosons, as
\begin{eqnarray}
\quad \quad\quad \quad\quad \quad\quad \quad AAW &:&  q \bar{q}^\prime \to H^\pm (\to A W) A \to 4b + X\,, \nonumber \\
AAAW &:& q \bar{q}^\prime \to H^\pm (H^\pm \to A W) H (\to A A) \to 4b+X\,, \nonumber \\  
AAZW &:& q \bar{q}^\prime \to H^\pm (H^\pm \to A W) H (\to A Z) \to 4b+X\,.\nonumber 
\end{eqnarray} 
We assume that the $W$ and $Z$ decay inclusively (i.e., both hadronically and leptonically), and $X$ can be additional quarks (including $b$-quarks) and/or leptons.

Fixing $m_A$ to two representative values of 50 and 70 GeV, and $m_h$ to 125 GeV, we scanned the remaining free parameters of the model in the ranges
\begin{center}
$m_H$: $[2m_A-250]$\,GeV\,,~~$m_{H^\pm}$: [100 -- 300]\,GeV\,,\\
$s_{\beta - \alpha}$: $0.9$ -- 1.0\,,~~$m_{12}^2$: 0 -- $m_A^2\sin\beta\cos\beta$\,,~~$t_\beta$: 1 -- 60\,, \\
\end{center}
to find their combinations that give substantial production cross sections for the processes noted above. The constraints imposed on the parameters during their numerical scanning include those from (1) theory, i.e., perturbativity, vacuum stability, and unitarity, (2) measurements of EW precision observables parameterized in terms of the oblique parameters ($S,~T$, and $U$), (3) measurement of BR$B\to X_s \gamma$, and (4) the 95$\%$ confidence limits from the direct Higgs boson searches at LEP, Tevatron, and LHC, and also from the Higgs precision measurements. 

From the successfully scanned parameter space, we chose two benchmark points, BP1 and BP2, given in Tab.~\ref{Table: BP1 and BP2}. BP1 corresponds to $m_A=70$ GeV, and therefore has the $h\to AA$ decay unavailable, the BR($H\to AA$) almost $1$, and the BR$(H\to AZ)$ strongly suppressed. For BP2, wherein $m_A=50$ GeV, the BR$H\to AZ=0.5$, while the $h\to AA$ decay is also open, and can thus give us an upper bound on the $h-A-A$ trilinear coupling (see Ref.~\cite{Sanyal:2023pfs}). $m_H$ and $m_{H^\pm}$ are each almost identical for both the BPs, as it allows us to assess the impact of the variation in $m_A$ alone on our analysis.

\begin{table}[h]
\centering
\resizebox{\textwidth}{!}{%
\begin{tabular}{|c|c|c|c|c|c|c|c|c|}
\hline
BP & $m_A$ [GeV] & $m_{H^\pm}$ [GeV] & $m_{H}$ [GeV] & $\tan\beta$ & $\sin(\beta - \alpha)$ & $m_{12}^2$ [GeV$^2$] & BR($H \to AA$) & BR($H \to AZ$) \\
\hline
1 & 70 & 169.7 & 144.7 & 7.47 & 0.988 & 2355.0 & 0.99 & 0.006 \\
2 & 50 & 169.8 & 150.0 & 17.11 & 0.975 & 1275.0 & 0.48 & 0.505 \\
\hline
\end{tabular}
}
\caption{Input parameter values, and BRs of the $H$ for the two selected BPs.}
\label{Table: BP1 and BP2}
\end{table}

\vspace{-5mm}
\section{Signal Isolation}

The backgrounds for the multi-$b$ final state include mult-jet and $t\bar{t}+$jets dominantly. In our computation, we match the multi-jet background up to four jets, and the $t\bar{t}$ up to two jets. The cross section for the multi-jet background in the $5$-flavor scheme at $\sqrt{s}=13$ TeV LHC is $9\times 10^6$ pb with the {\tt NNPDF23\_lo\_as\_0130}~\cite{NNPDF:2017mvq} parton distribution function (PDF), and for $t\bar{t}+$jets it is $834$ pb. For the two selected BPs, we estimate the multi-$b$ signal events with the same PDF set. The cross sections corresponding to the various signal modes, assuming a next-to-next-to-leading order $k$-factor of 1.35 \cite{Bahl:2021str}, are given in Tab.~\ref{Table: EW cross sections}. We perform event generation and parton shower with {\tt MadGraph5\_aMC@NLO}~\cite{Alwall:2014hca} and 
{\tt Pythia-8.2}~\cite{Sjostrand:2014zea}, using the anti-$k_t$ algorithm \cite{Cacciari:2008gp} with $R=0.4$ for jet-reconstruction. The detector simulation is performed using the {\tt Delphes-3.4.2} \cite{deFavereau:2013fsa} program. For $b$-tagging, we use the $p_T$-dependent efficiencies corresponding to the `DeepCSV Medium' working point based on the $\sqrt{s}=13$\,TeV data from the CMS collaboration \cite{CMS:2017wtu}. The primary selection cuts we apply on the jets to reconstruct all three BSM Higgs bosons include $p_T > 20$\,GeV and $|\eta| < 2.5$. Further selections that we make for each of the Higgs states are explained below.

{
\begin{table}[h]
\centering
\resizebox{\textwidth}{!}{%
\tiny\tiny
\begin{tabular}{|c|c|c|c|c|c|c|c|c|}
\hline
BP & $AAA$ [fb] & $AAZ$ [fb] & $AAWW$ [fb] & $AAW$ [fb] & $AAAW$ [fb] & $AAZW$ [fb]\\
\hline
1  & 171.6 & 0.76 & 25.2 & 142.3 & 79.7 & 0.35\\

2  & 101.3 & 79.3 & 27.7 & 198.0 & 37.1 & 29.0\\
\hline
\end{tabular}}
\caption{EW production cross sections for each of the signal channels for the two BPs.}
\label{Table: EW cross sections}
\end{table}
} 

\subsection{Reconstruction of $A$}

\begin{enumerate} 
\item Since the signal processes contain at least two $A$s, each event should contain at least 4 $b$-jets, $a$, $b$, $c$ and $d$, which can be resolved into pairs 1 and 2. For this purpose we use a pairing algorithm for the leading 4 $b$-jets to choose that combination out of the possible three, ($a,b\,;\;c,d)$, $(a,c\,;\; b,d)$, and $(a,d\,;\; b,c)$, which minimizes 
\begin{align}
\Delta R = |\Delta R_1 - 0.8| + |\Delta R_2 - 0.8| \,.
\end{align}  
Here, for a given combination, 
\begin{eqnarray}
\Delta R_1 = \sqrt{(\eta_a - \eta_b)^2 + (\phi_a - \phi_b)^2}\,,
\quad \Delta R_2 = \sqrt{(\eta_c - \eta_d)^2 + (\phi_c - \phi_d)^2}\,,
\end{eqnarray}
and offsetting each of these by 0.8 omits the $b$-jet pairings with too large an overlap in the $\{\eta,\,\phi\}$ space. This algorithm is motivated by the idea that the $b$-jets coming from a resonance (presumably the $A$) are much closer together than the uncorrelated ones.

\item After the pairing, we impose the asymmetry cut,
\begin{equation}\label{eq:alpha}
\bar{\alpha} =  \frac{|m_1 - m_2|}{m_1 + m_2} < 0.2\,,
\end{equation}
where $m_1$ and $m_2$ are the invariant masses of the two selected $b$-jet pairs. This cut ensures that these pairs are from two identical resonances, i.e., the two $A$s. 
\end{enumerate}

\subsection{Reconstruction of $H^\pm$} 
\begin{enumerate}

\item Since the $q \bar{q}^\prime \to A_1 H^\pm \to A_1\,A_2\,W^\pm \to 4b+jj$ process also yields two $A$s, each event should contain at least 4 $b$-tagged jets, along with another pair of jets.

\item The invariant mass of the two leading jets should lie within the $m_{W} \pm 25$\,GeV mass window.

\item The four $b$-jets are combined into two $b$-jet pairs and only events in which the invariant mass of each of these pairs lie within a 45\,GeV window around $m_A$, while satisfying the asymmetry cut $\bar{\alpha} < 0.2$, are selected. The vector $p_T$-sum of the $b$-jet pairs should match the $p_T$ of the reconstructed $A$s, with $A_1$ and $A_2$ selected such that $p_T(A_1) > p_T(A_2)$ (implying that the softer $A_2$ originates from the $H^\pm$ decay). This criterion reduces the background significantly. 

\item $m_{H_\pm}$ is calculated as the invariant mass of the $2b+jj$ system, where `$2b$' is the softer pair from $A_2$. 

\item If the above conditions are satisfied by more than one pairings of the $b$-jets, the pair which maximizes the separation $\Delta R = \sqrt{(\Delta\eta)^2+(\Delta\phi)^2}$ of the reconstructed $H^\pm$ and $A_1$ is selected.
\end{enumerate}

\subsection{Reconstruction of $H$} \label{sec:h-rec}
\begin{enumerate}

\item The events corresponding to the signal process $q\bar{q} \to A_1 H \to A_1 A_2 A_3 \to 4b+X$ should contain at least six $b$-tagged jets, which are combined into three pairs, out of which the one with the invariant mass lying within a 45\,GeV window around $m_A$, and satisfying the $\bar{\alpha}$-cut is selected.

\item Summing the 4-momenta of the $b$-jets in a pair gives its $p_T$, and the pair with the highest $p_T$ our of the three is identified as the prompt $A_1$. The invariant mass of $H$ is reconstructed from the remaining 4$b$-jet system, expected to originate from the $H\to A_2A_3$ decay.

\item If multiple pairs of the $b$-jets satisfy the above criteria, we use the 4$b$-jet system that maximizes its separation from the third $b$-jet pair (from the prompt $A_1$) in the $\{\eta,\,\phi\}$ space, for reconstructing the $H$.

\item Due to limited (mis-)tagging efficiencies, the tagging of 6 $b$-jets is highly challenging. We therefore also use events with at least 5 $b$-jets for the $H$ reconstruction, and assume the light jet with the leading $p_T$ to be the 6th $b$-jet in the steps 1 -- 3 above. These steps are repeated sequentially for the jet with the next highest $p_T$ in cases where the leading jet fails to satisfy the pairing criteria, until the correct jet is found.  
\end{enumerate}

\section{Significances at the LHC}

\begin{figure}[h]
\centering
\includegraphics[width=4.275cm]{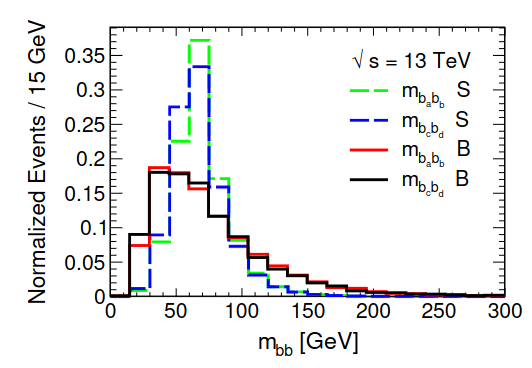}
\includegraphics[width=4.275cm]{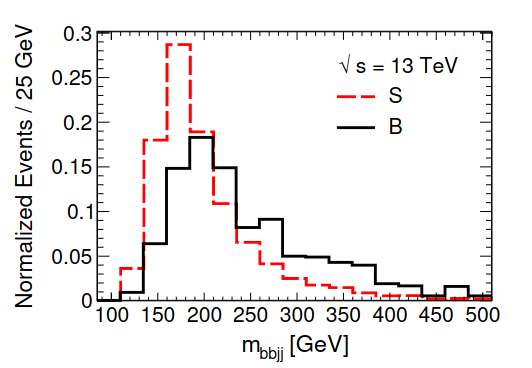}
\includegraphics[width=4.275cm]{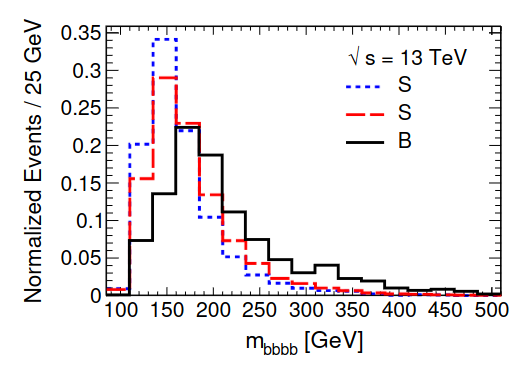}
\includegraphics[width=4.275cm]{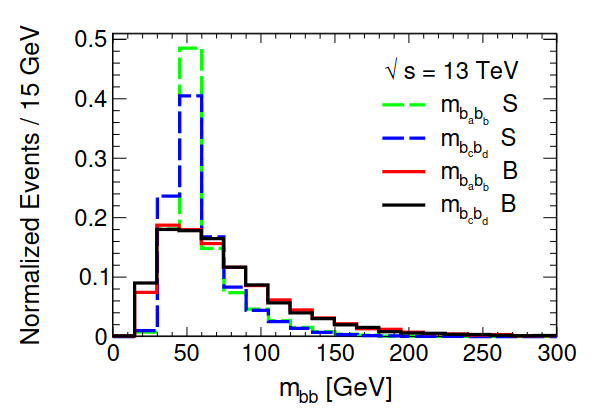}
\includegraphics[width=4.275cm]{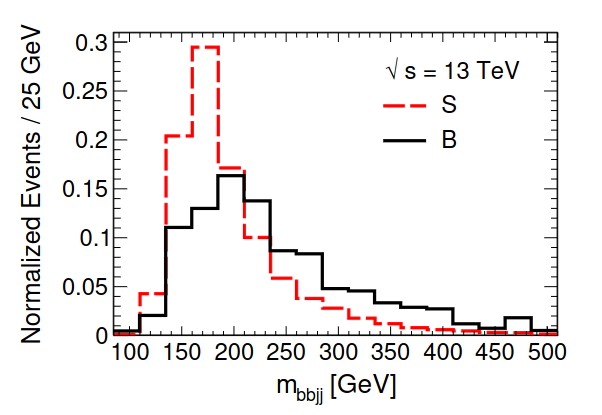}
\includegraphics[width=4.275cm]{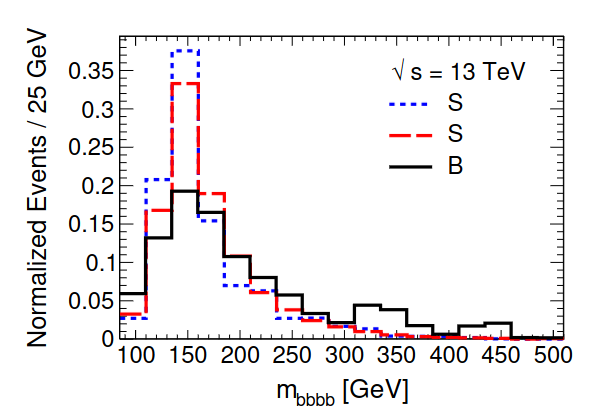}
\caption{The left frames show the $m_{bb}$ distributions for the signal (green/blue - dashed) and background (red/black - solid) events; the middle frames show the $m_{bbjj}$ distributions for the signal (red - dashed) and background (black - solid) events; and the right frames shows the $m_{bbbb}$ distributions for the signal (red/blue - dashed) and background (black - solid) events, for BP1 (top) and BP2 (bottom).}
\label{fig-1: Distributions}       
\end{figure}

We calculate the signal and background event rates, $S$ and $B$, respectively, for an integrated luminosity of 3000\,fb$^{-1}$ at the LHC. In Fig.~\ref{fig-1: Distributions} we show the normalized invariant mass distributions of the $b$-jet pairs. The top frame corresponds to BP1 and the bottom one to BP2. The subscript $a$ implies the distribution for the pair containing the leading $b$-jet. The signal distributions include all the EW modes, while both multi-jet and $t\bar{t}$+jets are included in the background distributions. The invariant masses peak at the true $m_A$ for both the BPs. Similarly, the $bbjj$ invariant mass distributions for the signal peak around the true $m_{H^\pm}$ for both the BPs.

The $m_{bbbb}$ distributions in Fig.~\ref{fig-1: Distributions} depicts the reconstruction of $H$, where the red-dashed signal histograms correspond to events with at least 5 $b$-jets, and peak around the true $m_H$ values. The blue-dotted histograms correspond to the invariant mass reconstructed using at least 6 $b$-tagged jets. While the latter provides better reconstruction, tagging 6 $b$-jets in the background events requires excessive computational resources, and we therefore settle for the 5$b$-jet events.

From these histrograms we choose three bins around the mass of each BSM Higgs boson to estimate the statistical significances of their signals,
which are given in Tab.~\ref{Table:significances}. Since the reconstruction of $A$ receives contributions from all the EW modes, the significance is highest for $A$ in both the BPs. The reconstruction of $H^\pm$ and $H$ are dependent on $AAW$ and $AAA$ modes only, and their significances of their signals are thus slightly lower. In the case of BP2, the lighter mass of $A$ results in much softer $b$-jets, which reduces the significance of $A$ reconstruction compared to BP1. Similarly, the $H$ signal significance is also reduced due to a lower BR$(H \to AA)$. Nevertheless, the $>$ 3$\sigma$ signal significances obtained for all the three Higgs bosons in BP2 are still formidable, demonstrating the strength of our proposed reconstruction methods. 
 
\begin{table}[h]
\small
\begin{tabular}{|c|ccc|ccc|ccc|}
\hline
& \multicolumn{3}{c|}{$A$} & \multicolumn{3}{c|}{$H^\pm$} & \multicolumn{3}{c|}{$H$} \\ \hline
~BP~ & \multicolumn{1}{c|}{~$\sigma_S$ [fb]~} & \multicolumn{1}{c|}{~$\sigma_B$ [fb]~} & ~$\frac{S}{\sqrt{B}}$~ & \multicolumn{1}{c|}{~$\sigma_S$ [fb]~} & \multicolumn{1}{c|}{~$\sigma_B$ [fb]~} & ~$\frac{S}{\sqrt{B}}$~ & \multicolumn{1}{c|}{~$\sigma_S$ [fb]~} & \multicolumn{1}{c|}{~$\sigma_B$ [fb]~} & ~$\frac{S}{\sqrt{B}}$~ \\ \hline
1 & \multicolumn{1}{c|}{15.4} & \multicolumn{1}{c|}{8864} & 8.9$\sigma$ & \multicolumn{1}{c|}{2.22} & \multicolumn{1}{c|}{482} & 5.5$\sigma$ & \multicolumn{1}{c|}{2.55} & \multicolumn{1}{c|}{309} & 7.9$\sigma$ \\
2 & \multicolumn{1}{c|}{10.4} & \multicolumn{1}{c|}{10175} & 5.7$\sigma$ & \multicolumn{1}{c|}{1.33} & \multicolumn{1}{c|}{491} & 3.3$\sigma$ & \multicolumn{1}{c|}{1.06} & \multicolumn{1}{c|}{256} & 3.6$\sigma$ \\ \hline
\end{tabular}
\caption{Total signal and background cross sections after applying all the selection cuts, and the discovery significances for the three BSM Higgs bosons.}
\label{Table:significances}
\end{table}

\section{Prospects at the $e^+ e^-$ colliders}

The electron-positron colliders anticipated in the near future, including the International Linear Collider (ILC) \cite{ILCInternationalDevelopmentTeam:2022izu}, the Future Circular Collider (FCC-ee) \cite{FCC:2018evy}, and the Circular Electron-Positron Collider (CEPC) \cite{CEPCStudyGroup:2018ghi} can serve as the most efficient machines to study the EW processes discussed in our work. Since the electron and positron are elementary particles, as opposed to composite protons, their collisions result in much cleaner backgrounds in comparison to the hadron colliders. Furthermore, the well-defined energy of the initial state (again, unlike at the hadron colliders), which coincides with $\sqrt s$ (up to some initial state radiation effects), could greatly improve the kinematic reconstruction of the final state. 

From the perspective of the Type-I 2HDM, when all the new Higgs bosons can be relatively light ($\mathcal{O}(100)$ GeV), the $e^+ e^-$ colliders, functioning as Higgs-factories initially with $\sqrt{s}=250$ GeV, will serve as great resources for probing them with upgrades to 500 GeV or higher energies. However, in $e^+ e^-$ colliders, the charged  two-body Higgs states are forbidden, and consequently the $AAW$ mode used to reconstruct $H^\pm$ is not viable. This challenge can still be addressed by applying a modification of our $H^\pm$ reconstruction algorithm to the charge-neutral mode
\begin{align}
AAWW ~~~: ~~~ e^+ e^- \to H^+ (\to AW) H^- (\to AW)   \to 4b + 4j\,. 
\end{align}

For this channel, events with at least 4 $b$-jets and 4 light jets will need to be tagged, followed by their grouping into two $2b2j$ systems. The correct pair should exhibit minimal asymmetry, $\bar{\alpha}$, (see Eq.~\ref{eq:alpha}), and as a result, it should reconstruct $H^\pm$. We plan to conduct such a study as a follow up to the current one. Another aspect we aim to explore is the advantage of polarized beams in linear colliders like the ILC, which could enhance the cross sections for, and hence the sensitivity to, the EW processes of our interest. To this end, we present the cross sections of the EW processes at $\sqrt{s}=250$ GeV and 500 GeV for our selected BPs with unpolarized beams in Tab.~\ref{ILC 250} and Tab.~\ref{ILC 500}, which are relevant for all prospective $e^+ e^-$ colliders.  

\begin{table}[t]
\centering
\begin{tabular}{|c|c|c|c|c|c|c|c|c|}
\hline
BP & $AAA$ [fb] & $AAZ$ [fb] & $AAWW$ [fb]\\
\hline
1 & 34.94 & 0.15 & $-$ \\

2 & 24.61 & 19.32 & $-$ \\
\hline
\end{tabular}
\caption{EW production cross sections at an $e^+e^-$ collider with $\sqrt{s}=250$ GeV in each of the signal channels for our two selected BPs. The $AAWW$ mode is kinematically inaccessible, and is therefore indicated with a "$-$" symbol.  
}
\label{ILC 250}
\end{table}

\begin{table}[t]
\centering
\begin{tabular}{|c|c|c|c|c|c|c|c|c|}
\hline
BP & $AAA$ [fb] & $AAZ$ [fb] & $AAWW$ [fb]\\
\hline
1 & 42.96 & 0.18 & 39.92 \\

2 & 21.48 & 16.86 & 43.77 \\
\hline
\end{tabular}
\caption{EW production cross sections at an $e^+e^-$ collider with $\sqrt{s}=500$ GeV in each of the signal channels for our two selected BPs.}
\label{ILC 500}
\end{table}

\section{Conclusions}

For understanding the origin of EWSB via the Higgs mechanism, a full reconstruction of all the Higgs states in a given BSM framework is required. However, any additional Higgs bosons, besides the one with a mass of 125 GeV discovered at the LHC in 2012, have remained elusive in the multitude of dedicated searches to date. In our view, the reason for their non-discovery hitherto is that these searches primarily focus on QCD-induced production of the additional Higgs states to assess their discovery prospects. In certain BSM frameworks, the couplings of the new Higgs bosons to SM fermions can be significantly suppressed, rendering them fermiophobic, and hence inaccessible in the conventional probes. The EW production of such Higgs bosons can dominate over QCD production, as we have established for the Type-I 2HDM in our study. 

We have demonstrated that all the three non-SM Higgs bosons in the Type-I 2HDM can be detected at the LHC in one unique final state containing 4 or more $b$-jets, which originate from EW processes. The proposed $e^+e^-$ colliders offer an ideal platform for not only the detection but also a much more clean and precise mass-reconstruction of these additional Higgs bosons, thanks to their much cleaner background environments compared to the hadron colliders. 

\section*{Acknowledgments}
P.S. would like to thank the organizers of the 2024 International Workshop on Future Linear Colliders (LCWS2024) for the opportunity to present this work. T.M. is supported by BITS Pilani Grant NFSG/PIL/2023/P3801. S.Mo. is supported in part through the NExT Institute and the STFC Consolidated Grant No. ST/L000296/1.  

\bibliography{LCWS_2HDM}

\begin{thebibliography}{18}

\bibitem{Mondal:2023wib}
T.~Mondal, S.~Moretti, S.~Munir, P.~Sanyal, {Electroweak Multi-Higgs
  Production: A Smoking Gun for the Type-I Two-Higgs-Doublet Model}, Phys. Rev.
  Lett. \textbf{131}, 231801 (2023), \texttt{2304.07719}.
  \doiwoc{10.1103/PhysRevLett.131.231801}

\bibitem{Enberg:2018pye}
R.~Enberg, W.~Klemm, S.~Moretti, S.~Munir, {Electroweak production of multiple
  (pseudo)scalars in the 2HDM}, Eur. Phys. J. C \textbf{79}, 512 (2019),
  \texttt{1812.01147}. \doiwoc{10.1140/epjc/s10052-019-7025-8}

\bibitem{Mondal:2021bxa}
T.~Mondal, P.~Sanyal, {Same sign trilepton as signature of charged Higgs in two
  Higgs doublet model}, JHEP \textbf{05}, 040 (2022), \texttt{2109.05682}.
  \doiwoc{10.1007/JHEP05(2022)040}

\bibitem{Kim:2022nmm}
J.~Kim, S.~Lee, J.~Song, P.~Sanyal, {Fermiophobic light Higgs boson in the
  type-I two-Higgs-doublet model}, Phys. Lett. B \textbf{834}, 137406 (2022),
  \texttt{2207.05104}. \doiwoc{10.1016/j.physletb.2022.137406}

\bibitem{Kim:2023lxc}
J.~Kim, S.~Lee, P.~Sanyal, J.~Song, D.~Wang, {$\tau^{ \pm} \nu \gamma \gamma$
  and $\ell^{ \pm} \ell^{ \pm} \gamma \gamma \rlap{\,/}{E}_T X$ to probe the
  fermiophobic Higgs boson with high cutoff scales}, JHEP \textbf{04}, 083
  (2023), \texttt{2302.05467}. \doiwoc{10.1007/JHEP04(2023)083}

\bibitem{Sanyal:2023pfs}
P.~Sanyal, D.~Wang, {Probing the electroweak $4b+\ell +\rlap{\,/}{E}_T$ final
  state in type I 2HDM at the LHC}, JHEP \textbf{09}, 076 (2023),
  \texttt{2305.00659}. \doiwoc{10.1007/JHEP09(2023)076}

\bibitem{Wang:2023pqx}
D.~Wang, J.H. Cho, J.~Kim, S.~Lee, P.~Sanyal, J.~Song, {Probing light
  fermiophobic Higgs boson via diphoton jets at the HL-LHC}, Phys. Rev. D
  \textbf{109}, 015017 (2024), \texttt{2310.17741}.
  \doiwoc{10.1103/PhysRevD.109.015017}

\bibitem{Branco:2011iw}
G.C. Branco, P.M. Ferreira, L.~Lavoura, M.N. Rebelo, M.~Sher, J.P. Silva,
  {Theory and phenomenology of two-Higgs-doublet models}, Phys. Rept.
  \textbf{516}, 1 (2012), \texttt{1106.0034}.
  \doiwoc{10.1016/j.physrep.2012.02.002}

\bibitem{NNPDF:2017mvq}
R.D. Ball et~al. (NNPDF), {Parton distributions from high-precision collider
  data}, Eur. Phys. J. C \textbf{77}, 663 (2017), \texttt{1706.00428}.
  \doiwoc{10.1140/epjc/s10052-017-5199-5}

\bibitem{Bahl:2021str}
H.~Bahl, T.~Stefaniak, J.~Wittbrodt, {The forgotten channels: charged Higgs
  boson decays to a W$^{±}$ and a non-SM-like Higgs boson}, JHEP \textbf{06},
  183 (2021), \texttt{2103.07484}. \doiwoc{10.1007/JHEP06(2021)183}

\bibitem{Alwall:2014hca}
J.~Alwall, R.~Frederix, S.~Frixione, V.~Hirschi, F.~Maltoni, O.~Mattelaer, H.S.
  Shao, T.~Stelzer, P.~Torrielli, M.~Zaro, {The automated computation of
  tree-level and next-to-leading order differential cross sections, and their
  matching to parton shower simulations}, JHEP \textbf{07}, 079 (2014),
  \texttt{1405.0301}. \doiwoc{10.1007/JHEP07(2014)079}

\bibitem{Sjostrand:2014zea}
T.~Sj\"ostrand, S.~Ask, J.R. Christiansen, R.~Corke, N.~Desai, P.~Ilten,
  S.~Mrenna, S.~Prestel, C.O. Rasmussen, P.Z. Skands, {An introduction to
  PYTHIA 8.2}, Comput. Phys. Commun. \textbf{191}, 159 (2015),
  \texttt{1410.3012}. \doiwoc{10.1016/j.cpc.2015.01.024}

\bibitem{Cacciari:2008gp}
M.~Cacciari, G.P. Salam, G.~Soyez, {The anti-$k_t$ jet clustering algorithm},
  JHEP \textbf{04}, 063 (2008), \texttt{0802.1189}.
  \doiwoc{10.1088/1126-6708/2008/04/063}

\bibitem{deFavereau:2013fsa}
J.~de~Favereau, C.~Delaere, P.~Demin, A.~Giammanco, V.~Lema\^\i{}tre,
  A.~Mertens, M.~Selvaggi (DELPHES 3), {DELPHES 3, A modular framework for fast
  simulation of a generic collider experiment}, JHEP \textbf{02}, 057 (2014),
  \texttt{1307.6346}. \doiwoc{10.1007/JHEP02(2014)057}

\bibitem{CMS:2017wtu}
A.M. Sirunyan et~al. (CMS), {Identification of heavy-flavour jets with the CMS
  detector in pp collisions at 13 TeV}, JINST \textbf{13}, P05011 (2018),
  \texttt{1712.07158}. \doiwoc{10.1088/1748-0221/13/05/P05011}

\bibitem{ILCInternationalDevelopmentTeam:2022izu}
A.~Aryshev et~al. (ILC International Development Team), {The International
  Linear Collider: Report to Snowmass 2021} (2022), \texttt{2203.07622}.

\bibitem{FCC:2018evy}
A.~Abada et~al. (FCC), {FCC-ee: The Lepton Collider}: {Future Circular Collider
  Conceptual Design Report Volume 2}, Eur. Phys. J. ST \textbf{228}, 261
  (2019). \doiwoc{10.1140/epjst/e2019-900045-4}

\bibitem{CEPCStudyGroup:2018ghi}
M.~Dong et~al. (CEPC Study Group), {CEPC Conceptual Design Report: Volume 2 -
  Physics \& Detector} (2018), \texttt{1811.10545}.

\end{thebibliography}
\end{document}